\documentclass[review]{elsarticle}

\usepackage{lineno,hyperref}
\usepackage{tikz}
\usepackage{tikz-cd}
\usepackage{dsfont}
\usepackage{amsfonts}
\usepackage{amsmath}
\usepackage{slashed}
\usepackage{graphics}
\usepackage{amsmath}
\usepackage{amssymb}
\usepackage[compat=1.1.0]{tikz-feynman}
\usepackage{setspace}
\usepackage{float}
\usepackage{cleveref}
\usepackage{units}
\usepackage{amscd}
\usepackage[left=1in, right=1in, top=1in, bottom=1in]{geometry}
\usepackage{nicefrac}
\usepackage{subcaption}
\modulolinenumbers[5]

\journal{IOP}
\begin{document}

\begin{frontmatter}

\title{\textbf{Testing the universality of quantum gravity theories with cosmic messengers in the context of DSR theories}}

\author{M.D.C. Torri$^{a*}$}

\cortext[mycorrespondingauthor]{M.D.C.Torri: marco.torri@unimi.it, marco.torri@mi.infn.it}

\address[Unimi]{Dipartimento di Fisica, Universit\'a degli Studi di Milano and INFN Milano\\via Celoria 16, I - 20133 Milano, Italy}

\begin{abstract}
Recently there have been several studies devoted to the investigation of the fate of fundamental relativistic symmetries at the foreseen unification of gravity and quantum regime, that is the Planck scale. In order to preserve covariance of the formulation even if in an amended formulation, new mathematical tools are required. In this work, we consider DSR theories that modify covariance by introducing a non-trivial structure in momentum space. Additionally, we explore the possibility of investigating both universal quantum gravity corrections and scenarios where different particle species are corrected differently within the framework of these models. Several astroparticle phenomena are then analyzed to test the phenomenological predictions of DSR models.
\end{abstract}


\end{frontmatter}

\section{Introduction}
In the attempt to formulate a complete quantum gravity (QG) theory, that can unify the quantum realm with General Relativity (GR), one of the most active research sectors concerns the fate of Lorentz symmetry at the Planck energy scale. Indeed, at this energy scale of about $\sim$$10^{19}$~{GeV}, the unification of the two physics realms is foreseen. Nowadays the Lorentz invariance is universally recognized as one of the most important symmetries underlying the formulation of the physical theories, indeed, it is  preserved in the formulation of GR and quantum mechanics (QM).

Until now it has been impossible to test any theory at the Planck energy scale, but in some theoretical scenarios, some departures from the Lorentz symmetry are predicted as residual signatures of the physics described by a more fundamental theory \cite{COST}. The first perturbative effect that can be investigated in the context of QG concerns the time delay that particles of different energies may accumulate during propagation. In this case, the QG perturbations are supposed equal for every particle species and influence the particle velocity that acquires a nontrivial dependence on the energy. Searching for QG phenomenological effects generated by such a fundamental and unified theory, one can probe the universality of the kinematics for different particle species. Therefore, evidence of a dependence of spacetime geodesics on the particle species can serve as a test of the validity of the weak equivalence principle (WEP). In this work we will consider an operational definition of the WEP, that is we will consider the universality of gravitational interaction. A more complete discussion about the various formulations of the equivalence principle can be found in \cite{Paunkovic}, where some motivations for possible violations are furnished and analyzed.

In the context of QG phenomenology investigation, one must distinguish between theories that predict a Lorentz invariance violation (LIV) such as the Standard Model Extension (SME) \cite{SME,SMECPT} from the theories that introduce a modification of this symmetry, for instance deforming the Poincarè algebra \cite{DSR,DSR2}, or modifying the momentum space \mbox{geometry \cite{HMSR}}, with a consequent deformation of the free particle kinematics.

In the case of SME, the search for new physics effects is not limited to QG, but it is set in a more general beyond the Standard Model (SM) research framework. The SME theory is formulated, including in the SM of particle physics, all the operators that preserve the usual gauge symmetry $SU(3)\times SU(2)\times U(1)$. The introduced perturbation operators are constructed as renormalizable or super-renormalizable in the minimal SME, or not renormalizable in the context of the non minimal SME. These perturbations are conceived as not universal and species-dependent in order to pose under test even the universality of the formulation of the physical theories. The SME is perhaps the most comprehensive research framework for conducting this kind of investigation, introducing a remarkably rich phenomenology that includes both high- and low-energy QG effects, while also exploring other exotic physics possibilities, such as CPT violation. The experimental results analyzed in the context of this research framework are used to pose constraints on the magnitude of the supposed LIV perturbations \cite{Russell}.

The DSR is conceived as a fundamental theory that introduces a modification of the Lorentz symmetry via a deformation of the fundamental symmetry algebra, that is the Poincarè one, affecting the free particle kinematics \cite{DSR,DSR2}.
In the context of DSR theories, there are some theoretical reasons suggesting the necessity for introducing different relativistic properties for different particle species \cite{Amelinop1,Amelinop2}, particularly in order to justify the different behavior of composed and more massive particles if compared with their elementary constituents. The induced phenomenology has not yet been deeply studied; besides only the demonstration of the possibility of formulating a fully relativistic, and not universal theory has been investigated.

Last but not least, { the Homogeneously Modified Special Relativity (HMSR) \cite{HMSR}} framework introduces a modification of the momentum space geometry that affects the free particle kinematics, and on the other hand, preserving a modified covariance formulation and deforming the Lorentz symmetry as in the DSR theories. In this framework, the QG-caused perturbations can acquire an explicit dependence on the particle species and this allows the introduction of a reach phenomenology. The violation of universality is a condition that underlies the possibility of introducing anomalous threshold energy effects, such as for the GZK cut-off effect \cite{Coleman,Stecker}, even in a theory that modifies but does not violate the covariance of the formulation \cite{Torri1,Torri2,Torri3}.

In this work, we will explore the possibility of searching QG signatures as departures from the predictions of the SM in the context of cosmic messengers \cite{COST,Whitepaper}, particularly neutrino physics and ultra high energy cosmic rays (UHECR).

Neutrino physics is, in general, an ideal playground for the search for new physics beyond the SM. Indeed, these particles furnish the first demonstrated example of physics beyond the SM (BSM). In the framework of the SM of particle physics, neutrinos are theoretically described as massless particles, but the well-established flavor oscillation phenomenon requires the presence of different neutrino mass eigenstates in order to occur. Therefore, neutrinos are interesting particles for searching for new physics effects and the most relevant BSM scenarios that can be tested in the neutrino sector are non-standard interactions (NSI) and QG perturbations. The proposed QG signature may manifest as modifications to the free particle kinematics, affecting the dispersion relations. In the context of QG perturbations common for every particle species, it is possible to detect the time delay accumulated by particles with varying energies thanks to the velocity dependence on the energy.

Furthermore, in a non-universal perturbation scenario, we demonstrate that modifications to the oscillation pattern, and the resulting survival probability, can be anticipated. In the following, we will explore the possibility of formulating some theoretical scenarios that can induce QG-caused effects in the free propagation of the different neutrino mass eigenstates. Indeed, if the formulated QG scenario foresees the possibility of non-universal QG kinematical corrections for different particle species, the neutrino propagation pattern can be affected by the modification of the survival probability.

In this work, we present a formulation of DSR theory that is fully relativistic, highlighting the non-universal character of the induced perturbations. We demonstrate that within the context of DSR theory, modifications to the interaction threshold energies are predicted while preserving the covariance of the theory, albeit in a modified form. In the following, we discuss the QG effects expected during astroparticle propagation, which result in time delays for particles of different energies.  Finally, we show that even the GZK phenomenon can be influenced by the perturbations predicted in non-universal DSR theories. Cosmic rays (CR) can be a useful framework to test the validity of the universality of quantum gravity-induced perturbations. The universe is opaque to the propagation of UHECR since they interact with the cosmic microwave background (CMB). Through this interaction process, named the GZK effect, a CR dissipates part of its energy. Different particle species are involved in the GZK phenomenon. { In fact, in the case of light CR protons, photopions and CMB radiation play a role. Non-universal QG perturbations can differently affect the species involved, altering this process and causing a dilation of the UHECR propagation path \cite{Stecker,Torri1,Torri2,Torri3}.} Searching for a modification of the cut-off predicted UHECR opacity sphere can be a candidate test for universality and the WEP.

\section{Introduction to Relative Locality Scenario and DSR Theories}
{Generally speaking, in some of the most studied modified relativity theories, particle kinematics is amended by the introduction of presumed QG perturbations \cite{COST}. The theoretical motivations for this approach are rooted in the idea that real physics takes place in phase space, where the interactions of different particles can be described. The main observations in physics concern the energies and momenta associated with particles, as well as the times of interactions. Spacetime, in contrast, emerges as a construct derived from the measurements made by local observers. Each observer builds their own spacetime as a local projection of momentum space, reflecting the fact that physical effects can only be detected in the vicinity of their measuring instruments.

This concept, referred to as \emph{Relative Locality} in the context of DSR theories \cite{DSR}, attributes presumed QG effects to a non-trivial curved phase space geometry, where the non-linear composition rules of momenta encode the geometric structure. The geometric properties of momentum space determine the modified kinematics of free-propagating or interacting particles. Because the composition rules in \emph{Relative Locality} are non-linear, the spacetime probed by particles explicitly depends on the energy of the probing particle. In contrast, \emph{Absolute Locality} corresponds to a flat momentum space with trivial geometry and \mbox{composition rules.} 

The ultimate objective of DSR models is to generalize Lorentz covariance without explicitly violating it. In these models, Lorentz symmetry is modified to account for the curved structure of momentum space with the associated non-trivial composition rules of momenta. In this context, the framework preserves generalized relativistic covariance, at least locally. At the same time, the modified symmetry must remain compatible with the standard formulation of Lorentz invariance in the low-energy limit, where QG effects are suppressed. This leads to a generalization of Special Relativity (SR), motivated by the broader attempt to unify quantum physics and gravity.

In the following, we will introduce the most studied \emph{Relative Locality} scenario, specifically focusing on the DSR theory based on the $\kappa$-Poincaré kinematic symmetry group \cite{Arzano}, and explore its implications for particle kinematics and symmetry.}

\section{Geometry of Curved Phase Space}
\label{sec3}
In this kind of DSR theory, the constructed momentum space is curved with a nontrivial geometrical structure. As proposed in \cite{Gubitosi,Amelino4} in the context of DSR theories, the modified on-shell relations can be defined using the geodesic distance constructed in a curved space, specifically the momentum space geodesic distance:
\begin{equation}
\label{dispersion}
\int_{0}^{1}\sqrt{g^{\mu\nu}(\gamma)\dot{\gamma}_{\mu}(s)\dot{\gamma}_{\nu}(s)}\,ds=d(0,p_{\mu})=m\qquad\gamma_{\mu}(0)=0,\;\gamma_{\mu}(1)=p_{\mu}
\end{equation}
where $g^{\mu\nu}(p)$ is the metric of the curved phase space. From \cref{dispersion} it is possible to obtain the geodesic equation:
\begin{equation}
\label{geodesicequat}
\frac{d^2\gamma_{\mu}}{d\tau^2}+\Gamma^{\alpha\beta}_{\mu}\frac{d\gamma_{\alpha}}{d\tau}\frac{d\gamma_{\beta}}{d\tau}=0
\end{equation}
where $\Gamma_{\mu}^{\alpha\beta}$ stands for the usual affine connection computed in the momentum space as follows:
\begin{equation}
\label{affineconn}
\Gamma_{\mu}^{\alpha\beta}=\frac{1}{2}g_{\mu\nu}\left(\frac{\partial\gamma^{\alpha\nu}}{\partial p_{\beta}}+\frac{\partial\gamma^{\nu\beta}}{\partial p_{\alpha}}-\frac{\partial\gamma_{\alpha\beta}}{\partial p_{\nu}}\right).
\end{equation}

Following the original geometric interpretation, proposed at first in \cite{Gubitosi,Amelino4} and then used in \cite{TorriNeutrino}, every point belonging to the curved momentum space $\mathcal{P}$ is connected to the origin of the space through a geodesic curve $\sigma(s):[0,1]\rightarrow\mathcal{P}$. The whole momentum space $\mathcal{P}$ is spanned by parametric surfaces $\sigma(s,t):[0,1]\times[0,1]\rightarrow \mathcal{P}$ defined such that for any couple of points $P,\,Q\in \mathcal{P}$ the geodesic curve $\sigma(s,0)$ is related to the first point $P$ and the curve $\sigma(0,t)$ is related to the second point $Q$. The vector $\frac{d\sigma(s,t)}{ds}$ is parallel transported along the tangent vector $\frac{d\sigma(s,t)}{dt}$. As a result, at any point of the parametric surface, after defining the covariant derivative associated with the affine connection \cref{affineconn}:
\begin{equation}
\nabla^{\mu}p_{\nu}=\dot{\partial}^{\mu}p_{\nu}+\Gamma^{\mu\alpha}_{\nu}p_{\alpha}
\end{equation}
where $\dot{\partial}^{\mu}=\frac{\partial}{\partial p_{\mu}}$ stands for the partial derivative computed with respect to the momentum.
The following parallel transport relation is satisfied:
\begin{equation}
\frac{d\sigma_{\mu}(s,t)}{dt}\nabla^{\mu}\frac{d\sigma_{\nu}(s,t)}{ds}=0.
\end{equation}

The previous relation can be used to construct a modified composition rule of momenta.
The modified composition rule can be introduced as the extremal point of the \mbox{parametric surface:}
\begin{equation}
p(P)\oplus q(Q)=\sigma(1,1).
\end{equation}

The construction is compatible with the usual definition of the modified composition rules used in DSR theories \cite{DSR} and is ruled by another connection whose definition is strictly related to translations:
\begin{equation}
\label{translationconn}
\widetilde{\Gamma}_{\mu}^{\alpha\beta}=-\frac{\partial}{\partial p_{\alpha}}\frac{\partial}{\partial p_{\beta}}\left(p\oplus_{k}q\right)_{\mu}\big|_{p=q=k=0}
\end{equation}
where the composition of momenta is translated to the point of coordinates $k$ using the opportune parametrical surface $\sigma^{k}(s,t)$.

As a consequence, the infinitesimal parallel transport relation for a momentum $p$, which is strictly related to the composition of momenta and to infinitesimal translations, can be obtained:
\begin{equation}
\left(p\oplus dq\right)_{\mu}=p_{\mu}+\tau_{\mu}^{\nu}(p)dq_{\nu}=p_{\mu}+q_{\mu}-\widetilde{\Gamma}_{\mu}^{\alpha\beta}p_{\alpha}q_{\beta}+\dots
\end{equation}
where the $\tau(p)$ term stands for the parallel transport coefficient.
The non-commutativity of the non-linear composition rules gives rise to torsion, which encodes the non-trivial geometric structure of momentum space:
\begin{equation}
-T^{\mu\nu}_{\alpha}=\frac{\partial}{\partial p_{\mu}}\frac{\partial}{\partial p_{\nu}}\left(p\otimes q-q\otimes p\right)_{\alpha}\bigg|_{p=q=0}.
\end{equation}

On the other hand, the non-associativity of the composition rules of momenta determines the momentum space curvature. In this case and related to the usual DSR formulation, the curvature is found to be identically zero:
\begin{equation}
R^{\mu\nu\alpha}_{\beta}=\frac{\partial}{\partial p_{[\mu}}\frac{\partial}{\partial q_{\nu]}}\frac{\partial}{\partial k_{\alpha}}\left[\left(\left(p\otimes q\right)\oplus k\right)-\left(p\oplus\left(q\oplus k\right)\right)\right]_{\beta}\bigg|_{p=q=k=0}=0.
\end{equation}

\section{The $\kappa$-Poincaré Algebra}
The construction of the symmetry group related to the DSR formulation is obtained in the context of Hopf algebras introducing a modification of the kinematics considering the $U(so(1,3))$ symmetry group together with the translation sector \cite{Majid,Lukierski,Lukierski2,Agostini}. In this work, we will consider the generalized $\kappa$-Poincaré structure introduced in \cite{Amelino4}; the commutators of the algebra commutators take the form
\begin{align}
\label{algebra}
&[P_{\mu},P_{\nu}]=0,\qquad [R_{i},R_{j}]=\epsilon_{ijk}R_{k},\qquad [N_{i},N_{j}]=-\epsilon_{ijk}R_{k}, \notag\\
&[R_{i},P_{0}]=0,\qquad [R_{i},P_{k}]=\epsilon_{ijk}P_{k},\qquad\;[R_{i},N_{j}]=\epsilon_{ijk}N_{k},\\
&[N_{i},P_{0}]=e^{\vartheta\lambda P_{0}}P_{i}, \notag\\&[N_{i},P_{j}]=\delta_{ij}\left(\frac{e^{(2-\vartheta)\lambda P_{0}}-e^{-\vartheta\lambda P_{0}}}{2\lambda}-\frac{\lambda}{2}e^{\vartheta\lambda P_{0}}|\vec{P}|^2\right)+(1-\vartheta)\lambda e^{\vartheta\lambda P_{0}}P_{i}P_{j}. \notag
\end{align}
here $P_{\mu}$, $N_{j}$, $R_{j}$ are the translation, boost and rotation generators. $\lambda$ is the usual deformation parameter $\lambda=1/\kappa\propto 1/M_{Pl}$, on the other hand, $\vartheta\in[0,\,1/2]$ parameterizes the different $\kappa$-Poincaré bases. The usual \emph{time to the right} $k$-Poincarè basis is obtained by choosing $\vartheta=0$, whereas the \emph{time symmetric} basis is derived for $\theta=1/2$.

The related Hopf algebra bicrossproduct structure is given by the following:
\begin{align}
\label{bicross}
&\Delta P_{0}=P_{0}\otimes\mathbb{I}+\mathbb{I}\otimes P_{0},\qquad \Delta P_{i}=P_{i}\otimes e^{-\vartheta\lambda P_{0}}+e^{(1-\vartheta)\lambda P_{0}}\otimes P_{i},\notag \\
&\Delta R_{i}=R_{i}\otimes\mathbb{I}+\mathbb{I}\otimes R_{i},\qquad \Delta N_{i}\otimes\mathbb{I}+e^{\lambda P_{0}}\otimes N_{i}-\lambda\epsilon_{ijk}e^{\vartheta\lambda P_{0}}P_{j}\otimes R_{k}.
\end{align}

The associated coalgebra antipodes $S$ and the counits $\epsilon$ associated with the generators $\{P_{\mu}\}$ are given by the following: 
\begin{align}
&S(P_{\mu}(p)=(\ominus p)_{\mu}\Rightarrow S(E)=-E,\;S(P)=-e^{(1-\vartheta)\lambda E}, \notag \\
&S(N)=-e^{(1-\vartheta)\lambda E}P, \notag\\ 
&P_{\mu}(0)=\epsilon(P_{\mu})\Rightarrow\epsilon(E)=\epsilon(P)=\epsilon(N)=0.
\end{align}

Classically in DSR theories, the dispersion relations are constructed starting from the Casimir operator of the algebra \cref{algebra}, named mass Casimir:
\begin{equation}
\label{MDR1}
\left(\frac{2}{\lambda}\right)^2\sinh^{2}\left(\frac{\lambda}{2}P_{0}\right)-|\vec{P}|^2e^{2(1-\vartheta)\lambda P_{0}}-m^2=0.
\end{equation}

In this work we will consider the geodesic distance as the dispersion relation, as illustrated in {\cref{sec3}}, and we will demonstrate that the two definitions coincide at least at the leading order of the perturbative series in $\lambda$.

\section{Specializing the Geometry}
\label{sec6}
The modified composition rules of momenta determine the geometric structure of momentum space. As a consequence, the connection \cref{translationconn}, related to the composition of momenta \cite{Gubitosi, Amelino4}, can be computed as follows:
\begin{equation}
\left(p\oplus q\right)_{0}=p_{0}+q_{0},\qquad\qquad \left(p\oplus q\right)_{j}=p_{j}e^{\vartheta\lambda q_{0}}+e^{(1-\vartheta)\lambda p_{0}}q_{0}. 
\end{equation}

Moreover, the induced translations in coordinate space are given by the following:
\begin{equation}
\label{translationconn2}
\widetilde{\Gamma}_{\mu}^{\alpha\beta}(p)=\delta_{\mu}^{\,j}\left(\vartheta\lambda\delta^{\alpha}_{\,0}\delta^{\beta}_{\,j}+(1-\vartheta)\lambda\delta^{\alpha}_{\,j}\delta^{\beta}_{\,0}\right).
\end{equation}

To characterize the momentum space and construct the related phase space it can be useful to determine the Killing vectors via the usual Killing equation:
\begin{equation}
\nabla^{\mu}\xi^{\nu}+\nabla^{\nu}\xi^{\mu}=0
\end{equation}
where the covariant derivative is defined using the affine connection \cref{affineconn} associated with the metric used in the MDR definition: $\nabla^{\mu}\xi^{\nu}=\dot{\partial}^{\mu}\xi^{\nu}+\Gamma^{\mu\nu}_{\vartheta}\xi^{\vartheta}$.
The solutions are \mbox{given by the following:}
\begin{equation}
\xi^{\nu}_{\mu}(p)=\left(
                \begin{array}{cccc}
                  1 & -\vartheta\lambda p_{1}& -\vartheta\lambda p_{2} & -\vartheta\lambda p_{3} \\
                  0 & e^{(1-\vartheta)\lambda p_{0}} & 0 & 0 \\
                  0 & 0 & e^{(1-\vartheta)\lambda p_{0}} & 0 \\
                  0 & 0 & 0 & e^{(1-\vartheta)\lambda p_{0}} \\
                \end{array}
              \right).
\end{equation}

The Killing vectors can be used to define the vierbein related to the geometry of the momentum space $e^{\mu}_{\,\nu}(p)=\xi^{\mu}_{\,\nu}(p)$. Using this vierbein, the metric associated with the momentum space can be computed by obtaining the following:
\begin{align}
ds^2=&e^{\mu}_{\,\alpha}(p)\eta^{\alpha\beta}e^{\nu}_{\,\beta}(p)dp_{\mu}dp_{\nu}=\left(1-(\vartheta\lambda p)^2\right)dp_{0}^{2}+2\vartheta\lambda e^{(1-\vartheta)\lambda p_{0}}\vec{p}\,d\vec{p}\,dp_{0} \notag \\
-&e^{2(1-\vartheta)\lambda p_{0}}d\vec{p}^2.
\end{align}

This result is compatible with that obtained from the time-ordered plane waves derived from different choices of momentum space bases \cite{Amelino4}.

The coordinate space can be constructed, for instance, starting from the vectors $\{\chi^{\mu}\}$ defined such that they satisfy the canonical Poisson brackets together with the \mbox{momenta $\{p_{\mu}\}$:}
\begin{equation}
\{\chi^{\mu},\chi^{\nu}\}=0,\qquad\{p_{\mu},p_{\nu}\},\qquad\{\chi^{\mu},p_{\nu}\}=\delta^{\mu}_{\nu}.
\end{equation}

Moreover, the phase space coordinates can be constructed using the Killing \mbox{vectors \cite{Amelino4}:}
\begin{equation}
x^{\mu}=e^{\mu}_{\nu}(p)\chi^{\nu}=\xi_{\nu}^{\mu}(p)\chi^{\nu}.
\end{equation}

The resulting modified Poisson brackets between coordinates are as follows:
\begin{equation}
\{x^{0},x^{j}\}=\lambda x^{j},\qquad \{x^{i},x^{j}\}=0,
\end{equation}
demonstrating the non-commutativity of the obtained spacetime.

The remaining Poisson brackets are written as follows:
\begin{align}
&\qquad\{p_{\mu},p_{\nu}\}=0,\qquad \{p_{0},x^{j}\}=0,\\
&\{p_{j},x^{0}\}=(1-\vartheta)\lambda p_{j}, \qquad \{p_{j},x^{k}\}=\delta_{j}^{k}. \notag
\end{align}

The resulting spacetime is no longer commutative and the commutators of the coordinates are as follows:
\begin{equation}
[x^{\mu},\,x^{\nu}]=i\zeta^{\mu\nu}_{\alpha}x^{\alpha}
\begin{cases}
\zeta^{\mu\nu}_{\alpha}=(1-\vartheta)\;\text{for}\,\mu=0,\,\nu\in\{1,\,2,\,3\}\\
\zeta^{\mu\nu}_{\alpha}=0\;\text{for the other index combinations}.
\end{cases}
\end{equation} 

\section{DSR Action Including Boundary Translation Terms}
Using the previous results on the formulation of spacetime coordinates, we can address the construction of the action for both free-propagating and interacting particles. The explicit action can be formulated as follows \cite{DSR}:
\begin{equation}
\label{action}
S=\sum_{j\in J}\int d\tau\bigg\{-x_{(j)}^{\mu}\dot{p}_{\mu(j)}+\mathcal{N}_{j}C_{j}(p,\,m)+\zeta^{\mu}\mathcal{K}_{\mu}\left(p_{1}(\tau),\ldots,p_{n}(\tau)\right)\bigg\}
\end{equation}
where $C_{j}(p,\,m)$ is the dispersion relation obtained from the geodesic distance \cref{dispersion}:
\begin{equation}
C_{j}(p,\,m)=d(0,\,p)^2-m^2
\end{equation}
and $\mathcal{N}$ is a Lagrange multiplier enforcing the MDR in the action. Furthermore, the $\zeta$ are the Lagrangian multipliers enforcing the preservation of translations, whose generators $\mathcal{K}^{j}$ encode the modified composition rules of momenta. From the variation of the action \cref{action}, the equations of motion can be obtained:
\begin{align}
&\frac{dx^{\mu}_{(j)}}{dt}=\mathcal{N}_{j}\frac{\partial d(0,\,p)}{\partial p_{\mu(j)}},\\
&\frac{dp_{\mu(j)}}{dt}=0.
\end{align}

Also, the MDR and modified composition rules of momenta are obtained from the variation in the action:
\begin{align}
&d^{2}(0,\,p)-m^2=0,\\
&\mathcal{K}_{\mu}(p_{1},\dots,p_{n})=0.
\end{align}

In this context, the coordinates of the interaction points can be computed, obtaining a result compatible with the construction of the spacetime coordinates presented in \cref{sec6}:
\begin{equation}
x^{\mu}_{(j)}(0)=\zeta^{\nu}\frac{\partial}{\partial p_{\mu(j)}}\mathcal{K}_{\nu}(p_{1},\dots,p_{n}).
\end{equation}

Consequently, all interaction worldlines converge at a single spacetime point, and the non-trivial geometric structure of momentum space gives rise to the concept of \emph{Relative Locality} \cite{DSR}.

The translations are generated using the Poisson brackets following the prescription:
\begin{equation}
\label{tralgen}
\delta x^{\mu}_{(j)}=\epsilon^{\nu}\{\mathcal{K}_{\nu},x^{\mu}_{(j)}\}=\epsilon^{\mu}-\epsilon^{\alpha}\widetilde{\Gamma}_{\alpha}^{\mu\nu}p_{\nu(j)}.
\end{equation}

Therefore, the non-trivial geometric structure of the momentum space affects the translations that acquire an explicit dependence on the probe energy, an explicit manifestation of the \emph{Relative Locality} idea.

\section{Covariance of the Formulation}
Since we are dealing with a DSR theory, we have to check that the formulation preserves an amended form of covariance. It is straightforward to check that the on-shell relation covariance is preserved in \cref{dispersion}. Indeed, the formulation of the geodesic distance is invariant under the action of momentum space passive diffeomorphisms. For instance, since the action of a diffeomorphism $\widetilde{p}=f(p)$ on momentum space tangent vectors is given by the following:
\begin{equation}
\widetilde{\dot{p}}=\frac{\partial f(p)_{\alpha}}{\partial p_{\nu}}\dot{p}_{\nu}
\end{equation}
and this transformation relation is valid for the curves defined in the momentum space {
\begin{equation}
\label{comrulinv}
\widetilde{\dot{\gamma}}(\tau)=\frac{\partial f(\gamma(\tau))_{\alpha}}{\partial \gamma_{\nu}(\tau)}\dot{\gamma}_{\nu}(\tau),
\end{equation}}
the diffeomorphism action on the metric is resumed by the relation
\begin{equation}
\widetilde{g}^{\mu\nu}(f(p))=\frac{\partial f(p)_{\alpha}}{\partial p_{\mu}}g^{\alpha\beta}(p)\frac{\partial f(p)_{\beta}}{\partial p_{\nu}}.
\end{equation}

As a result, the geodesic distance defining the on-shell relation \cref{dispersion} results is covariant and the geodesic \cref{geodesicequat} preserves the covariance of the formulation.

An issue is caused by the presence in the action \cref{action} of the translation generators \cref{tralgen} and the necessity to guarantee the covariance of these terms. This problem is extensively treated in \cite{Amelino4} demonstrating that the translation generators can be deformed by the action of diffeomorphisms, such that the resulting theory is not compatible with the starting one. This point imposes some restrictions on the possibility to preserve covariance and particular caution must be posed in obtaining modifications to the theory under the action of coordinate transformations.

Now the invariance of the modified composition rules of momenta is checked. This aspect of DSR theories is extensively analyzed in literature \cite{Gubitosi,Amelino4}. The DSR-amended Lorentz transformations can be constructed requiring the condition of compatibility with the modified composition rules of momenta:
\begin{equation}
\label{crc}
\Lambda\left(p\oplus q\right)=\Lambda p \oplus \Lambda q.
\end{equation}

In order to preserve the composition rule covariance \cref{crc}, the action of the Lorentz transformations must include the so-called backreaction \cite{Gubitosi,Bruno,Amelinotrasf}; indeed, if $\beta$ is the Lorentz coefficient associated with the Lorentz transformation, the action on the composition rule becomes the following:
\begin{equation}
\Lambda(\beta)\left(p\oplus q\right)=\Lambda(\beta)p\oplus\Lambda(\beta\triangleleft p) q
\end{equation}
and $\beta\triangleleft p$ stands for the backreaction of the momentum $p$ applied to the coefficient $\beta$ for the second transformation.

At this stage, it is important to underline that if it is possible to define symmetry transformations that preserve the invariance in the composition rules of momenta \cref{comrulinv}, a momentum space connection can be defined through \cref{translationconn}.

\section{Modified Dispersion Relation and Time Delay}
As previously mentioned, the MDR can be constructed starting from the Casimir operator of the modified algebra \cref{algebra}, leading to the explicit form given in \cref{MDR1}. In this work, we consider the MDR \cref{dispersion} constructed using the geodesic distance \cite{Amelino4}, resulting in an expression that depends on the choice of basis, encoded by the coefficient $\vartheta \in [0,\,1/2]$:{
\begin{equation}
\label{MDR2}
\frac{1}{\lambda}\operatorname{arcosh}\left(\cosh{(\lambda p_{0})} -\frac{\lambda^{2}}{2}e^{2\lambda(1-\vartheta)p_{0}}|\vec{p}|^2\right)-m^{2}=0.
\end{equation}}

Both forms of the MDR \cref{MDR1,MDR2} can be approximated by the \mbox{following expression:}
\begin{equation}
\label{MDR}
p_{0}^{2}-e^{2\lambda(1-\vartheta)p_{0}}|\vec{p}|^{2}-m^2\simeq p_{0}^{2}-(1+2\lambda(1-\vartheta)p_{0})|\vec{p}|^{2}.
\end{equation}

The particle energy can be computed starting from the obtained MDR:
\begin{equation}
E\simeq\sqrt{|\vec{p}|^{2}\left(1+\lambda(1-\vartheta)p_{0}\right)}\simeq\sqrt{|\vec{p}|^{2}\left(1+\delta/M_{Pl}(1-\vartheta)|\vec{p}|\right)}.
\end{equation}

In the last equality, the relation $\lambda = \delta / M_{Pl}$ is used, where $\delta$ encodes the magnitude of the QG perturbation relative to the normalization factor given by the Planck mass $M_{Pl}$. Applying the Hamilton equation to the computed energy, the particle velocity can then \mbox{be determined:}
\begin{equation}
\vec{v}(E)=\frac{\partial}{\partial \vec{p}}\;p_{0}\simeq \frac{\vec{p}+3/2\cdot\delta\cdot \vec{p}\cdot|\vec{p}|/M_{Pl}}{\sqrt{|\vec{p}|^{2}\left(1+\delta/M_{Pl}(1-\vartheta)|\vec{p}|\right)}}.
\end{equation}

The velocity explicitly depends on the particle energy, and its functional form is determined by the choice of the $\delta$ parameter \cite{TorriNeutrino}. 
The intensity of QG effects depends on the choice of basis in momentum space. Specifically, with the \emph{time to the right} basis ($\vartheta = 0$), the effect is maximal, while its magnitude is minimized when the \emph{time symmetric} basis is used ($\vartheta = 1/2$). As pointed out in \cite{Amelino4}, the description of physics depends on the choice of basis in momentum space. In the case of a positive $\delta$, superluminal neutrinos could arise; however, this possibility is excluded, as such particles would lose energy during propagation through pair production \cite{Vissani, Vissani2}. Conversely, a negative $\delta$ ensures subluminal neutrinos, with their velocity decreasing as energy increases. Thanks to this phenomenon, particles accelerated with different energies can present a different time arrival, accumulating time delay during their propagation. This effect can be investigated in the context of astroparticles \cite{TorriNeutrino}, which provide the ideal playground for this search due to their high energy and long propagation paths, allowing for the accumulation of the presumed tiny QG perturbations. The time delay can be easily computed in the context of the DSR framework, obtaining the following:
\begin{equation}
\Delta t=\frac{|\delta|}{M_{PL}} E(z)=\frac{|\delta|}{M_{PL}}\int_{0}^{z}\frac{(1+\zeta)E}{H_{0}\sqrt{\Omega_{\Lambda}+\Omega_{m}(1+\zeta)^3}}\,d\zeta
\end{equation}
where $E(z)$ is the energy depending on the redshift coefficient $z$ related to the particles source. $H_{0}$, $\Omega_{\Lambda}$ and $\Omega_{m}$ denote the Hubble constant, the cosmological parameter and the matter fraction, respectively.

In the astroparticle sector, gamma-ray bursts (GRBs) represent a suitable class of phenomena for investigating this effect since they emit nearly simultaneous photons and possibly neutrinos \cite{Rachen, Guetta, Ahlers, IceCube2014, Kimura, Kimura2} with a range of energies. The prompt neutrino emission phase in GRBs is not yet well understood; nevertheless, some studies predict the feasibility of conducting this research. The results obtained so far appear promising, as there is good agreement between observed data and theoretical predictions \cite{Amelino1,Amelino2,Ellis,GuettaPiran,Jacob}.

Another promising sector for investigating time delay is the detection of neutrinos accelerated during a stellar collapse, followed by a supernova (SN) explosion \cite{Mirizzi,Janka}. As discussed below, particles with varying energies may exhibit different times of flight from the source to the detector. The associated time delay can be observed by analyzing the emitted neutrino energy spectrum. A more comprehensive discussion on this topic can be found in \cite{TorriNeutrino} and the references therein.

The scenario investigated in this section is based on the hypothesis of universal QG perturbations, which affect all particle species in the same way. This demonstrates the possibility of identifying physical phenomena that can be influenced by QG perturbations without violating the universality of the interaction. Instead, in the following, we will address the possibility of studying non-universal QG perturbations predicted in the context of DSR theories and, as a consequence, the WEP.

\section{Mixing Algebras in Particle Depending Theories}
At this stage, we will introduce the idea necessary to ensure the possibility of investigating nonuniversal QG modifications in the context of DSR theories. In the following, we will illustrate how it can be possible to combine the Hopf algebras related to different spaces probed by different particle species.

In this work, we generalize the construction made in \cite{Amelinop1,Amelinop2,TorriNeutrino} by considering the dependence on the choice of basis in momentum space.
First, we present a method for combining the algebras, followed by the definition of the corresponding mixed coproduct. Then, we derive how to compose the four-momenta of different particle species within the Hopf algebra framework. In this context, the application of the mixing of the coproduct must be defined as the mathematical structure that generates the modified composition rules for momenta from distinct algebras. For example, the coproduct of elements from different algebras $H_{j}$ can be defined by introducing a support algebra $H'$ , which is associated with a projection map for each algebra $\phi_{j}$:
\begin{equation}
\phi_{j}:H_{j}\rightarrow H'.
\end{equation}

Below, we provide a concrete example of how the projection map can be defined for the $\kappa$-Poincaré algebras underlying the symmetry groups of different particles. The projection map $\phi$ can be constructed by relating the bicrossproduct basis generators of the various algebras as follows:
\begin{equation}
\phi(P_{\mu})=\frac{\lambda'}{\lambda}P_{\mu}',\quad\phi(R_{i})=R'_{i},\quad\phi(N_{i})=N'_{i}
\end{equation}
where $\{P_{\mu}\},\,\{R_{i}\},\,\{N_{i}\}$ are the four-momenta, the rotations and the boosts, respectively, while $\lambda$ denotes the characteristic coefficients of the different $\kappa$-Poincaré algebras.\\
Using the previous definitions it is now straightforward to obtain the relations:
\begin{align}
\label{a0}
&[\phi(P_{\mu}),\,\phi(P_{\nu})]=\left(\frac{\lambda'}{\lambda}\right)^{2}[P'_{\mu},\,P'_{\nu}]=0=\phi([P_{\mu},\,P_{\nu}]),\notag\\
&[\phi(R_{i}),\,\phi(R_{j})]=[R'_{i},\,R'_{j}]=\epsilon_{ijk}R'_{k}=\phi([R_{i},\,R_{j}]),\notag\\
&[\phi(N_{i}),\,\phi(N_{j})]=[N'_{i},\,N'_{j}]=-\epsilon_{ijk}R'_{k}=\phi([N_{i},\,N_{j}]),\notag\\
&[\phi(R_{i}),\,\phi(N_{j})]=[R'_{i},\,N'_{j}]=\epsilon_{ijk}N'_{k}=\phi([N_{i},\,P_{j}]),\notag\\
&[\phi(R_{i}),\,\phi(P_{0})]=\frac{\lambda'}{\lambda}[R'_{i},\,P'_{0}]=0=\phi([R_{i},\,P_{0}]).
\end{align}

Finally, the following relations can be verified by applying the previous results:{
\begin{align}
\label{a1}
&[\phi(R_{i}),\,\phi(P_{j})]=\frac{\lambda'}{\lambda}[R'_{i},\,P'_{j}]=\epsilon_{ijk}\frac{\lambda'}{\lambda}P'_{k}=\phi([R_{i},\,\phi(P_{j})]),\notag\\
&[\phi(N_{i}),\,\phi(P_{0})]=\frac{\lambda'}{\lambda}e^{\theta\lambda'P'_{0}}P'_{i}=\frac{\lambda'}{\lambda}[N'_{i},\,P'_{0}]=\phi([N_{i},\,P_{0}])
\end{align}}
and finally the relation
\begin{align}
\label{a2}
&[\phi(N_{i}),\,\phi(P_{j})]=\notag\\
&=\frac{\lambda'}{\lambda}\delta_{ij}\left(\frac{e^{(2-\theta)\lambda'P'_{0}}-e^{-\theta\lambda'P'_{0}}}{2\lambda'}-\frac{\lambda'}{2}e^{\theta\lambda'P'_{0}}|\vec{P}'|^2\right)+(1-\theta)\lambda'e^{\theta\lambda'P'_{0}}P'_{i}P'_{j}=\\\notag
&=\delta_{ij}\left(\frac{e^{(2-\theta)\lambda'P'_{0}}-e^{-\theta\lambda'P'_{0}}}{2\lambda}-\frac{\lambda}{2}e^{\theta\lambda'P'_{0}}|\vec{P}'|^2\right)+(1-\theta)\lambda e^{\theta\lambda'P'_{0}}P'_{i}P'_{j}=\phi([N_{i},\,P_{j}]).
\end{align}

As a consequence  of the previous \cref{a0,a1,a2} it is possible to state that $H'$ is a Hopf algebra since the projection map preserves the commutation rules. Since the previous relations are valid for every choice of algebra basis parameterized by the coefficient $\vartheta$, this statement is independent of the used basis, as is expected.

The mixed coproduct can be defined by combining the projection maps associated with different algebras. For example, in the case of two distinct Hopf algebras, $H_{1}$ and $H_{2}$, the corresponding projection maps $\phi_{1}$ and $\phi_{2}$ are as follows:
\begin{align}
\label{mixcoprod}
&\qquad\Delta':H'\otimes H'=\phi_{1}(H_{1})\otimes\phi_{2}(H_{2})\notag\\
&\begin{CD}
H'@>\Delta'>>H'\otimes H'@>\phi_{1}^{-1}\otimes\phi_{2}^{-1}>>H_{1}\otimes H_{2}.
\end{CD}
\end{align}

In the coalgebra sector, the following relations are valid for the coproduct:{
\begin{align}
&\Delta'(P'_{0})=\Delta'(\phi(P_{0}))=\frac{\lambda'}{\lambda}P'_{0}\otimes\mathbb{I}+\mathbb{I}\otimes\frac{\lambda'}{\lambda}P'_{0}=\phi\otimes\phi(\Delta(P_{0})),\notag\\
&\Delta'(P'_{i})=\Delta'(\phi(P_{i}))=\frac{\lambda'}{\lambda}\left(P'_{i}\otimes e^{-\theta\lambda'P'_{0}}+e^{(1-\theta)\lambda'P'_{0}}\mathbb{I}\otimes P'_{i}\right),\notag\\
&\Delta'(R'_{i})=\Delta'(\phi(R_{i}))=R_{i}\otimes\mathbb{I}+\mathbb{I}\otimes R_{i},\notag\\
&\Delta'(N'_{i})=\Delta'(\phi(N_{i}))=N'_{i}\otimes\mathbb{I}+e^{\lambda'P'_{0}}\otimes N'_{i}-\frac{\lambda'}{\lambda}\epsilon_{ijk}e^{\theta\lambda'P'_{0}}P'_{j}\otimes R'_{k}.
\end{align}}

To prove that the $\phi$ map is an isomorphism, we must verify its compatibility with the antipodes and the counits. We check the compatibility with the antipodes using the following relations:

\begin{align}
&\phi(S(P_{0}))=-\frac{\lambda'}{\lambda}P'_{0}=S'(\phi(P_{0})),\notag\\
&\phi(S(P_{i}))=-\frac{\lambda'}{\lambda}e^{-\lambda'P'_{0}}P'_{i},\notag\\
&\phi(S(R_{i}))=-R_{i}=S'(\phi(R_{i})),\notag\\
&\phi(S(N_{i}))=-e^{\lambda'P'_{0}}N'_{i}+\frac{\lambda'}{\lambda}\epsilon_{ijk}e^{\lambda'P'_{0}}P'_{j}R'_{k}=S'(\phi(N_{i})),
\end{align}
where $S'$ represents the antipode map related to the algebra $H'$. The compatibility with the co-unit is straightforward. The definition of the inverse map $\phi^{-1}:H'\rightarrow H$ is simply obtained in the following form:
\begin{equation}
\phi^{-1}(P'_{\mu})=\frac{\lambda}{\lambda'}P_{\mu},\quad\phi^{-1}(R'_{i})=R_{i},\quad\phi^{-1}(N'_{i})=N_{i}.
\end{equation}

Now it is possible to state that $\phi:H\rightarrow H'$ is an isomorphism relating the $\kappa$-Poincaré algebras $H$ and $H'$.

As a final result, now we can introduce the momentum-modified composition rules starting from the \cref{mixcoprod}:
\begin{align}
&p_{0}\oplus'_{\lambda_{1}\lambda_{2}}q_{0}=\frac{\lambda'}{\lambda_{1}}p_{0}+\frac{\lambda'}{\lambda_{2}}q_{0},\notag\\
&p_{i}\oplus'_{\lambda_{1}\lambda_{2}}q_{i}=\frac{\lambda'}{\lambda_{1}}e^{\theta q_{0}}p_{i}+\frac{\lambda'}{\lambda_{2}}e^{(1-\theta)\lambda' p_{0}}q_{i}.
\end{align}

By exchanging the deformation coefficients $\lambda_{1},\,\lambda_{2}$ and $\lambda'$, the inverted maps and the reversed order composition rules can be obtained.

\section{Neutrino Oscillations}
The possibility of mixing algebras associated with different particle species opens up opportunities to investigate QG perturbations that lack a universal character \cite{TorriNeutrino}. The phenomenon of neutrino oscillation represents an ideal area for this type of \mbox{investigation \cite{Torri4,Torri5,Torri6,Krauss,Gasperini,Mann}}, as different QG perturbations can affect the mass eigenstates involved in neutrino propagation differently, resulting in a modified oscillation pattern.

The neutrino oscillation is ruled by the Schro\"edinger equation and the solution is expressed using the particle momentum. Starting from the MDR \cref{MDR} it is possible to obtain the following relation:
\begin{equation}
|\vec{p}_{j}|\simeq\left(1-\frac{\delta}{M_{Pl}}E\right)E+\frac{m_{j}^2}{2E}
\end{equation}
valid for every $j$ mass eigenstate. The oscillation phase related to the mass eigenstate can be computed as follows:
\begin{equation}
\phi_{j}\simeq\left(\frac{\delta_{j}}{M_{Pl}}E^{2}-\frac{m_{j}^{2}}{2E}\right)L
\end{equation}
where the perturbation coefficient $\delta_{j}$ is related to the $j$-esim mass eigenstate $m_{j}$. The difference between two phases related to two different mass eigenstates is as follows:
\begin{equation}
\Delta\phi_{jk}=\left(\frac{\Delta m_{jk}^{2}}{2E}-\frac{\delta_{jk}}{M_{Pl}}E^{2}\right)L,
\end{equation}
where {
\begin{align}
&\Delta m_{jk}^{2}=m_{j}^{2}-m_{k}^{2} \notag\\
&\delta_{jk}=\delta_{j}-\delta_{k}.
\end{align}}

The QG perturbation effect becomes noticeable if $\delta_{jk}$ is nonzero, meaning the correction is not universal and varies for each particle species.

\begin{figure}[H]
\begin{centering}
\includegraphics[scale=0.45]{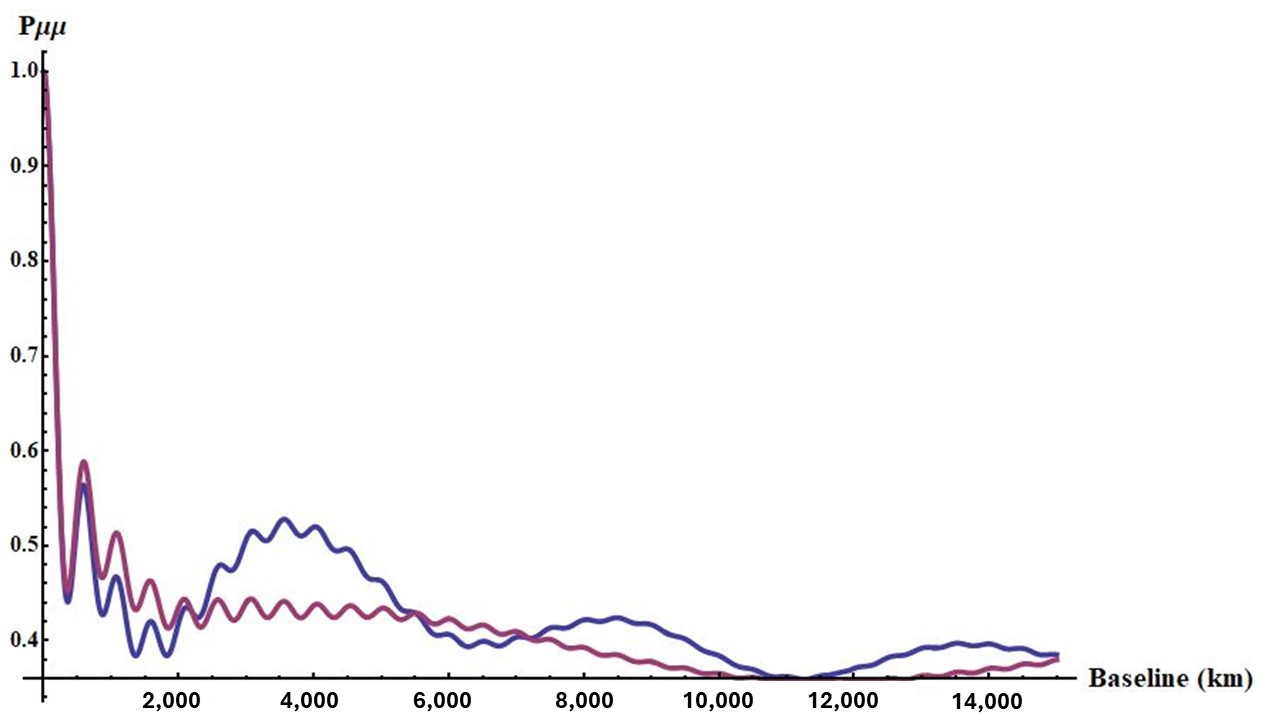}
\caption{Comparison of the atmospheric neutrino survival probability, integrated over the energy range: 500 Mev--5 GeV, as a function of the baseline: standard case (blue line) vs HMSR (red line).}
\label{fig1}
\end{centering}
\end{figure}

The impact of a nonuniversal correction in the oscillation phenomenon can be detected in the atmospheric sector comparing the expected flux of different neutrino flavors with the detected one. Here, we report the plot of the integrated survival probability of muonic neutrinos \cref{fig1} within the energy range of $500$ MeV - $5$ GeV:
\begin{equation}
P_{\mu\mu}=\frac{\int_{E_{min}}^{E_{max}}\phi_{\nu}(E)P(\nu_{\mu}\rightarrow\nu_{\mu})(E)dE}{\int_{E_{min}}^{E_{max}}\phi_{\nu}(E)dE}
\end{equation}
where $P(\nu_{\mu}\rightarrow\nu_{\mu})(E)$ is the survival probability and $\phi_{\nu}(E)$ is the expected neutrino flux as a function of energy.

\section{Threshold Reaction Energy Modifications}
The modification induced by QG can affect the threshold energy required for certain physical phenomena to occur. Indeed, the proposed QG perturbations can impact particle kinematics, introducing modifications such as the MDR. These kinematic perturbations are encoded in the non-trivial geometric structure of the momentum space, which leads to differences in the computations carried out within it. The threshold energy of a physical phenomenon is related to the free energy or Mandelstam $s$ variable. In the case of a two-particle interaction, the free energy depends on the internal product defined in the momentum space as follows:
\begin{equation}
\label{mandelstam}
s=\langle p+q|p+q\rangle
\end{equation}
where $p$ and $q$ are the four-momenta related to the particles interacting.  The internal product of \cref{mandelstam} can be defined so that its results are compatible with the MDR \cref{MDR}:
\begin{align}
\label{mandelstam1}
s=&\langle p+q|p+q\rangle= \notag\\
=&p_{0}^{2}-|\vec{p}|^{2}\left(1+2\lambda'(1-\vartheta)p_{0}\right)+q_{0}^{2}-|\vec{q}|^{2}\left(1+2\lambda'(1-\vartheta)q_{0}\right)+2p_{0}q_{0} \notag\\
-&2\vec{p}\cdot\vec{q}\left(1+2\lambda'(1-\vartheta)\sqrt{p_{0}q_{0}}\right)
\end{align}
where $\lambda'$ is the parameter related to the support Hopf algebra, used to describe the different particle interactions. This definition of free energy is covariant according to the definition used in DSR theory. In fact, the Mandelstam $s$ is defined starting from the MDR, which is invariant by construction. Therefore, the threshold energy associated with physical processes can be impacted by the introduction of DSR-predicted perturbations, which preserve Lorentz covariance, even if in an amended formulation.

\section{Impact on the Cosmic Rays Propagation}
A first example of a physical process whose threshold energy can be impacted by DSR predictions is the Greisen–Zatsepin–Kuzmin (GZK) effect \cite{Stecker,Torri1,Torri2,Torri3}, thereby influencing the CR propagation. CR can be categorized as light particles, such as protons, and heavy ones, such as bare nuclei. The Universe is opaque to the propagation of ultra-high-energy cosmic rays (UHECR). During their propagation, UHECR can interact with the CMB, dissipating energy. This energy dissipation occurs through pair production, photodissociation, and, for the most energetic particles, the GZK effect. For example, protons with energy exceeding a threshold of E$\sim$5 $\times10^{15}$ eV can interact with the CMB, resulting in the production of a $\Delta$ particle resonance. This process generates a photopion and reduces the energy of the initial CR proton:
\begin{equation}
p+\gamma\rightarrow\Delta\rightarrow 
\begin{cases}
p+\pi^{0}\\
n+\pi^{+}.
\end{cases}
\end{equation}

In order to guarantee the possibility of the $\Delta$ resonance necessary for the GZK phenomenon, the free energy must satisfy the following relation:
\begin{equation}
\label{b1}
s=\langle p_{p}+p_{\gamma}|p_{p}+p_{\gamma}\rangle \geq m_{\Delta}^{2}.
\end{equation}

Using the definition \cref{mandelstam1} in \cref{b1}, it is possible to obtain the following:
\begin{align}
\label{GZK}
m_{p}^{2}-&2(\lambda'-\lambda_{p})(1-\vartheta)E_{p}|\vec{p}_{p}|^{2}-2(\lambda'-\lambda_{\gamma})E_{\gamma}|\vec{p}_{\gamma}|^{2}+2E_{p}E_{\gamma}\notag \\
-&2\vec{p}_{p}\cdot\vec{p}_{\gamma}\left(1+2\lambda'(1-\vartheta)\sqrt{E_{p}E_{\gamma}}\right)\geq m_{\Delta}^{2}
\end{align}
where $\lambda_{p},\,\lambda_{\gamma}$ are the QG parameters associated with the proton and the photon, respectively, and $\lambda'$ is associated with the algebra used to set the interaction. In the previous relation, the MDRs of the proton and the photon were used. In this case, as well, the predicted physical effects are found to depend on the choice of basis in momentum space. By performing the computation in a reference frame where the photon energy is negligible compared to the UHECR proton energy ($E_{\gamma}\ll E_{p}$), one can derive the following constraint from \cref{GZK}:
\begin{equation}
\lambda'-\lambda_{p}<\frac{E_{p}}{2(1-\vartheta)|\vec{p}|^{2}}
\end{equation}

The difference between the correction to the proton and the correction related to the algebra used to describe the interaction should be limited to $\lambda'-\lambda_{p}<10^{-20}$ eV in order to ensure that the GZK effect is not suppressed.

The magnitude of the QG correction can be evaluated considering the attenuation length associated with the GZK effect. The attenuation length of a CR is defined as the average distance a particle can travel through the CMB radiation before its energy { is reduced by a factor $1/e$}\footnote{$e$ is the Napier's number, the base of the natural logarithm}, due to interactions. In the context of UHECR protons the main dissipation mechanism is the GZK phenomenon and the inverse of the attenuation length can be defined as follows:
\begin{equation}
\frac{1}{l_{p\gamma}}=\int_{\epsilon_{thr}}^{+\infty}\int_{-\pi}^{\pi}n(\epsilon)\frac{1}{2}\,s\,(1-\mu)\,\sigma_{p\gamma}(s)\,K(s)\,d\epsilon\,d\cos{\theta}.
\end{equation}

Here, $\sigma_{p\gamma}(s)$ is the proton--photon interaction cross-section as a function of the Mandelstam variable, $s$, $n(\epsilon)$ is the photon background density distribution and $\cos{\theta}$ is the impact parameter. $K(s)$ is the reaction inelasticity, defined as the fraction of energy available for secondary particle production during the reaction. The introduction of QG perturbations affects the inelasticity, leading to an enlargement of the predicted GZK opacity sphere. The computation is performed in the center-of-mass (CM) reference frame for simplicity, assuming the creation of a pion from a proton. This reference frame is defined via the relation $\vec{p}^{\,*}_{p}+\vec{p}^{\,*}_{\pi}=0$ where { the index $*$ denotes the physical quantities defined in the CM reference frame.} The four momentum associated with the pion in the CM can be expressed as follows: $p_{\pi}^{*}=(E_{\pi}^{\,*},\,\vec{p}^{\,*}_{\pi})=(\sqrt{s}-E_{p}^{*},\,\vec{p}_{p}^{\,*})$. The energy required to produce a photopion can be computed using the MDR as follows:
\begin{equation}
\label{b2}
s=\langle p_{\pi}|p_{\pi}\rangle=E_{\pi}^{2}-(1+2\lambda_{\pi}(1-\vartheta)E_{\pi})=m_{\pi}^{2}.
\end{equation}

From \cref{b2} and the definition of momenta in the CM frame, the following equation, valid in the CM reference frame, can be derived:
\begin{equation}
E_{p}^{*}=F(s)=\frac{s+m_{p}^{2}-m_{\pi}^{2}-2(\lambda_{\pi}-\lambda_{p})(1-\vartheta)|\vec{p}^{\,*}_{p}|^{2}E^{*}_{p}}{2\sqrt{s}}.
\end{equation}

The high-energy limit, where $E^{}_{p} \simeq |\vec{p}^{,}_{p}|$, QG introduces a perturbation that increases the predicted residual energy of the proton after the interaction, provided the condition $\lambda{p} > \lambda_{\pi}$ is satisfied. The final computation of the inelasticity in the high-energy limit gives
\begin{equation}
\label{b3}
(1-K_{\pi}(\theta))=\frac{1}{\sqrt{s}}\left(F(s)+\cos{\theta}\sqrt{F(s)^{2}-m_{p}^{2}+2(\lambda_{p}-\lambda_{\pi})(1-\vartheta)|\vec{p}|^{2}E_{p}}\right).
\end{equation}

The final form of the inelasticity is obtained by averaging over the direction
\begin{equation}
\label{b4}
K_{\pi}=\frac{1}{\pi}\int_{0}^{\pi}K_{\pi}(\theta)\,d\theta.
\end{equation}

From \cref{b3,b4}, it is straightforward to deduce that the inelasticity increases compared to the standard physics scenario. Furthermore, if the proton's correction is greater than that associated with the pion, the GZK opacity sphere expands.

\section{Conclusions}
In this research work, we analyzed the phenomenological predictions of DSR theories, presenting several astroparticle scenarios in which to test them. Initially, we introduced the models of DSR theories along with the corresponding modifications to the algebra of symmetry group generators, which are identified as $\kappa$-Poincaré. We illustrated how to construct the mathematical foundation of the theory and how it can be set within the framework of Hopf algebras. Starting from the modified momentum composition laws, the non-trivial geometry of momentum space was constructed. We demonstrated how these modifications impact particle kinematics, altering the dispersion relations. The MDRs were introduced as the Casimir operator of the modified algebra, as well as through the geodesic distance defined in momentum space. These modifications were derived in the context of a universal framework, assuming that all particles are equally affected by the introduction of QG. We then analyzed the first phenomenological effect arising from universal kinematic modifications, showing that particle velocities acquire an energy dependence. As a result, higher-energy particles can accumulate a time delay compared to lower-energy particles. This effect can be observed in the context of neutrino physics related to SN and GRB.

Next, we demonstrated that for DSR theories defined within the framework of Hopf algebras, it is possible to consider non-universal scenarios of QG-induced perturbations. Specifically, it is feasible to associate different corrected algebras with distinct particles, using a common algebra to define the interaction between different species via a projection. Two areas were considered to test this presumed non-universality of QG perturbations. The first area of investigation involves atmospheric neutrinos, where different corrections for the various mass eigenstates involved in propagation can impact the neutrino oscillation pattern. Thanks to the high energies involved, atmospheric neutrinos provide an ideal framework for this research, allowing the accumulation of the small perturbations predicted by the theory.

Through the MDR, it was possible to define how to determine the free energy required for certain physical processes. It was shown that threshold energy modifications remain invariant under the modified Lorentz covariance because they are constructed using the MDRs, which are invariant by design. This result should be considered a generalization of what was demonstrated in \cite{AmelinoFate}. In that work, it was shown that it is not possible to preserve covariance while modifying threshold energies for physical effects in universal scenarios. However, by considering non-universal corrections, we have demonstrated that it is possible to preserve the covariance defined for DSR theories while modifying threshold energies. We then showed that a sector where this research can be conducted is the GZK effect for UHECRs. The modifications induced by DSR models can expand the predicted GZK opacity sphere, altering the kinematics and the resulting inelasticity of processes such as photopion production.

As a final observation, all the analyzed scenarios were considered by introducing a dependence on the choice of basis in momentum space. The predicted effects may vary in intensity depending on the chosen basis. However, they are not completely suppressed. This behavior aligns with the predictions made in \cite{Amelino4}.


\begin{thebibliography}{999}
\bibitem{COST}
A.~Addazi, J.~Alvarez-Muniz, R.~Alves Batista, G.~Amelino-Camelia, V.~Antonelli, M.~Arzano, M.~Asorey, J.~L.~Atteia, S.~Bahamonde and F.~Bajardi, \textit{et al.},
``Quantum gravity phenomenology at the dawn of the multi-messenger era\textemdash{}A review,''
Prog. Part. Nucl. Phys. \textbf{125} (2022), 103948
doi:10.1016/j.ppnp.2022.103948
[arXiv:2111.05659 [hep-ph]].

\bibitem{Paunkovic}
N.~Paunkovic and M.~Vojinovic,
Universe \textbf{8} (2022) no.11, 598
doi:10.3390/universe8110598
[arXiv:2210.00133 [gr-qc]].

\bibitem{SME}
D.~Colladay and V.~A.~Kostelecky,
``Lorentz violating extension of the standard model,''
Phys. Rev. D \textbf{58} (1998), 116002
doi:10.1103/PhysRevD.58.116002
[arXiv:hep-ph/9809521 [hep-ph]].

\bibitem{SMECPT}
D.~Colladay and V.~A.~Kostelecky,
``CPT violation and the standard model,''
Phys. Rev. D \textbf{55} (1997), 6760-6774
doi:10.1103/PhysRevD.55.6760
[arXiv:hep-ph/9703464 [hep-ph]].

\bibitem{DSR}
G.~Amelino-Camelia, L.~Freidel, J.~Kowalski-Glikman and L.~Smolin,
``The principle of relative locality,''
Phys. Rev. D \textbf{84} (2011), 084010
doi:10.1103/PhysRevD.84.084010
[arXiv:1101.0931 [hep-th]].

\bibitem{DSR2}
G.~Amelino-Camelia,
``Doubly-Special Relativity: Facts, Myths and Some Key Open Issues,''
Symmetry \textbf{2} (2010), 230-271
doi:10.3390/sym2010230
[arXiv:1003.3942 [gr-qc]].

\bibitem{HMSR}
M.~D.~C.~Torri, V.~Antonelli and L.~Miramonti,
``Homogeneously Modified Special relativity (HMSR): A new possible way to introduce an isotropic Lorentz invariance violation in particle standard model,''
Eur. Phys. J. C \textbf{79} (2019) no.9, 808
doi:10.1140/epjc/s10052-019-7301-7
[arXiv:1906.05595 [hep-th]].

\bibitem{Russell}
V.~A.~Kostelecky and N.~Russell,
``Data Tables for Lorentz and CPT Violation,''
Rev. Mod. Phys. \textbf{83} (2011), 11-31
doi:10.1103/RevModPhys.83.11
[arXiv:0801.0287 [hep-ph]].

\bibitem{Amelinop1}
G.~Amelino-Camelia, M.~Palmisano, M.~Ronco and G.~D'Amico,
Int. J. Mod. Phys. D \textbf{29} (2020) no.02, 2050017
doi:10.1142/S0218271820500170
[arXiv:1910.05997 [gr-qc]].

\bibitem{Amelinop2}
G.~Amelino-Camelia,
Symmetry \textbf{4} (2012), 344-378
doi:10.3390/sym4030344
[arXiv:1111.5643 [hep-ph]].

\bibitem{Coleman}
S.~R.~Coleman and S.~L.~Glashow,
``High-energy tests of Lorentz invariance,''
Phys. Rev. D \textbf{59} (1999), 116008
doi:10.1103/PhysRevD.59.116008
[arXiv:hep-ph/9812418 [hep-ph]].

\bibitem{Stecker}
F.~W.~Stecker and S.~T.~Scully,
``Searching for New Physics with Ultrahigh Energy Cosmic Rays,''
New J. Phys. \textbf{11} (2009), 085003
doi:10.1088/1367-2630/11/8/085003
[arXiv:0906.1735 [astro-ph.HE]].

\bibitem{Torri1}
M.~D.~C.~Torri,
``Quantum Gravity Phenomenology Induced in the Propagation of UHECR, a Kinematical Solution in Finsler and Generalized Finsler Spacetime,''
Galaxies \textbf{9} (2021) no.4, 103
doi:10.3390/galaxies9040103
[arXiv:2110.09184 [gr-qc]].

\bibitem{Torri2}
M.~D.~C.~Torri, L.~Caccianiga, A.~di Matteo, A.~Maino and L.~Miramonti,
``Predictions of Ultra-High Energy Cosmic Ray Propagation in the Context of Homogeneously Modified Special Relativity,''
Symmetry \textbf{12} (2020) no.12, 1961
doi:10.3390/sym12121961
[arXiv:2110.09900 [hep-ph]].

\bibitem{Torri3}
M.~D.~C.~Torri, S.~Bertini, M.~Giammarchi and L.~Miramonti,
``Lorentz Invariance Violation effects on UHECR propagation: A geometrized approach,''
JHEAp \textbf{18} (2018), 5-14
doi:10.1016/j.jheap.2018.01.001
[arXiv:1906.06948 [hep-ph]].

\bibitem{Whitepaper}
R.~Alves Batista, G.~Amelino-Camelia, D.~Boncioli, J.~M.~Carmona, A.~di Matteo, G.~Gubitosi, I.~Lobo, N.~E.~Mavromatos, C.~Pfeifer and D.~Rubiera-Garcia, \textit{et al.}
``White Paper and Roadmap for Quantum Gravity Phenomenology in the Multi-Messenger Era'', Class. Quantum Grav. \textbf{42} (2025) 032001
doi:10.1088/1361-6382/ad605a 
[arXiv:2312.00409 [gr-qc]].

\bibitem{Arzano}
M.~Arzano and J.~Kowalski-Glikman,
``A group theoretic description of the \ensuremath{\kappa}-Poincar\'e Hopf algebra,''
Phys. Lett. B \textbf{835} (2022), 137535
doi:10.1016/j.physletb.2022.137535
[arXiv:2204.09394 [hep-th]].

\bibitem{Majid}
S.~Majid and H.~Ruegg,
``Bicrossproduct structure of kappa Poincare group and noncommutative geometry,''
Phys. Lett. B \textbf{334} (1994), 348-354
doi:10.1016/0370-2693(94)90699-8
[arXiv:hep-th/9405107 [hep-th]].

\bibitem{Lukierski}
J.~Lukierski, A.~Nowicki and H.~Ruegg,
``New quantum Poincare algebra and k deformed field theory,''
Phys. Lett. B \textbf{293} (1992), 344-352
doi:10.1016/0370-2693(92)90894-A

\bibitem{Lukierski2}
J.~Lukierski and H.~Ruegg,
``Quantum kappa Poincare in any dimension,''
Phys. Lett. B \textbf{329} (1994), 189-194
doi:10.1016/0370-2693(94)90759-5
[arXiv:hep-th/9310117 [hep-th]].

\bibitem{Agostini}
A.~Agostini, G.~Amelino-Camelia and F.~D'Andrea,
``Hopf algebra description of noncommutative space-time symmetries,''
Int. J. Mod. Phys. A \textbf{19} (2004), 5187-5220
doi:10.1142/S0217751X04020919
[arXiv:hep-th/0306013 [hep-th]].

\bibitem{Gubitosi}
G.~Gubitosi and F.~Mercati,
``Relative Locality in $\kappa$-Poincar\'e,''
Class. Quant. Grav. \textbf{30} (2013), 145002
doi:10.1088/0264-9381/30/14/145002
[arXiv:1106.5710 [gr-qc]].

\bibitem{Amelino4}
G.~Amelino-Camelia, S.~Bianco and G.~Rosati,
``Planck-Scale-Deformed Relativistic Symmetries and Diffeomorphisms on Momentum Space,''
Phys. Rev. D \textbf{101} (2020) no.2, 026018
doi:10.1103/PhysRevD.101.026018
[arXiv:1910.01673 [gr-qc]].

\bibitem{TorriNeutrino}
M.~D.~C.~Torri and L.~Miramonti,
Class. Quant. Grav. \textbf{41} (2024) no.15, 153001
doi:10.1088/1361-6382/ad5825
[arXiv:2404.04076 [gr-qc]].

\bibitem{Bruno}
N.~R.~Bruno, G.~Amelino-Camelia and J.~Kowalski-Glikman,
``Deformed boost transformations that saturate at the Planck scale,''
Phys. Lett. B \textbf{522} (2001), 133-138
doi:10.1016/S0370-2693(01)01264-3
[arXiv:hep-th/0107039 [hep-th]].

\bibitem{Amelinotrasf}
G.~Amelino-Camelia,
``Relativity in space-times with short distance structure governed by an observer independent (Planckian) length scale,''
Int. J. Mod. Phys. D \textbf{11} (2002), 35-60
doi:10.1142/S0218271802001330
[arXiv:gr-qc/0012051 [gr-qc]].

\bibitem{Vissani}
F.~L.~Villante and F.~Vissani,
``On the generality of the Cohen and Glashow constraints on the neutrino velocity,''
[arXiv:1110.4591 [hep-ph]].

\bibitem{Vissani2}
M.~Mannarelli, M.~Mitra, F.~L.~Villante and F.~Vissani,
``Non-Standard Neutrino Propagation and Pion Decay,''
JHEP \textbf{01} (2012), 136
doi:10.1007/JHEP01(2012)136
[arXiv:1112.0169 [hep-ph]].

\bibitem{Rachen}
J.~P.~Rachen and P.~Meszaros,
``Cosmic rays and neutrinos from gamma-ray bursts,''
AIP Conf. Proc. \textbf{428} (1998) no.1, 776-780
doi:10.1063/1.55402
[arXiv:astro-ph/9811266 [astro-ph]].

\bibitem{Guetta}
D.~Guetta, D.~Hooper, J.~Alvarez-Muniz, F.~Halzen and E.~Reuveni,
``Neutrinos from individual gamma-ray bursts in the BATSE catalog,''
Astropart. Phys. \textbf{20} (2004), 429-455
doi:10.1016/S0927-6505(03)00211-1
[arXiv:astro-ph/0302524 [astro-ph]].

\bibitem{Ahlers}
M.~Ahlers, M.~C.~Gonzalez-Garcia and F.~Halzen,
``GRBs on probation: testing the UHECR paradigm with IceCube,''
Astropart. Phys. \textbf{35} (2011), 87-94
doi:10.1016/j.astropartphys.2011.05.008
[arXiv:1103.3421 [astro-ph.HE]].

\bibitem{IceCube2014}
M.~G.~Aartsen \textit{et al.} [IceCube],
``Search for Prompt Neutrino Emission from Gamma-Ray Bursts with IceCube,''
Astrophys. J. Lett. \textbf{805} (2015) no.1, L5
doi:10.1088/2041-8205/805/1/L5
[arXiv:1412.6510 [astro-ph.HE]].

\bibitem{Kimura}
S.~S.~Kimura,
``Chapter 9: Neutrinos from Gamma-Ray Bursts,''
doi:10.1142/9789811282645\_0009
[arXiv:2202.06480 [astro-ph.HE]].

\bibitem{Kimura2}
K.~Murase, M.~Mukhopadhyay, A.~Kheirandish, S.~S.~Kimura and K.~Fang,
``Neutrinos from the Brightest Gamma-Ray Burst?,''
Astrophys. J. Lett. \textbf{941} (2022) no.1, L10
doi:10.3847/2041-8213/aca3ae
[arXiv:2210.15625 [astro-ph.HE]].

\bibitem{Amelino1}
G.~Amelino-Camelia, G.~D'Amico, G.~Rosati and N.~Loret,
``In-vacuo-dispersion features for GRB neutrinos and photons,''
Nature Astron. \textbf{1} (2017), 0139
doi:10.1038/s41550-017-0139
[arXiv:1612.02765 [astro-ph.HE]].

\bibitem{Amelino2}
G.~Amelino-Camelia, M.~G.~Di Luca, G.~Gubitosi, G.~Rosati and G.~D'Amico,
``Could quantum gravity slow down neutrinos?,''
Nature Astron. \textbf{7} (2023) no.8, 996-1001
doi:10.1038/s41550-023-01993-z
[arXiv:2209.13726 [gr-qc]].

\bibitem{Ellis}
J.~Ellis, N.~E.~Mavromatos, A.~S.~Sakharov and E.~K.~Sarkisyan-Grinbaum,
``Limits on Neutrino Lorentz Violation from Multimessenger Observations of TXS 0506+056,''
Phys. Lett. B \textbf{789} (2019), 352-355
doi:10.1016/j.physletb.2018.11.062
[arXiv:1807.05155 [astro-ph.HE]].

\bibitem{GuettaPiran}
G.~Amelino-Camelia, D.~Guetta, J.~Piran,
"ICECUBE Neutrinos and Lorentz Invariance Violation,"
Astrophys.J. \textbf{806} (2015) 2, 269
doi:10.1088/0004-637X/806/2/269

\bibitem{Jacob}
U.~Jacob and T.~Piran,
``Neutrinos from gamma-ray bursts as a tool to explore quantum-gravity-induced Lorentz violation,''
Nature Phys. \textbf{3} (2007), 87-90
doi:10.1038/nphys506
[arXiv:hep-ph/0607145 [hep-ph]].

\bibitem{Mirizzi}
A.~Mirizzi, I.~Tamborra, H.~T.~Janka, N.~Saviano, K.~Scholberg, R.~Bollig, L.~Hudepohl and S.~Chakraborty,
``Supernova Neutrinos: Production, Oscillations and Detection,''
Riv. Nuovo Cim. \textbf{39} (2016) no.1-2, 1-112
doi:10.1393/ncr/i2016-10120-8
[arXiv:1508.00785 [astro-ph.HE]].

\bibitem{Janka}
H.~T.~Janka,
``Neutrino-driven Explosions,''
doi:10.1007/978-3-319-21846-5\_109
[arXiv:1702.08825 [astro-ph.HE]].

\bibitem{Torri4}
M.~D.~C.~Torri,
``Neutrino Oscillations and Lorentz Invariance Violation,''
Universe \textbf{6} (2020) no.3, 37
doi:10.3390/universe6030037
[arXiv:2110.09186 [hep-ph]].

\bibitem{Torri5}
V.~Antonelli, L.~Miramonti and M.~D.~C.~Torri,
``Neutrino oscillations and Lorentz Invariance Violation in a Finslerian Geometrical model,''
Eur. Phys. J. C \textbf{78} (2018) no.8, 667
doi:10.1140/epjc/s10052-018-6124-2
[arXiv:1803.08570 [hep-ph]].

\bibitem{Torri6}
V.~Antonelli, L.~Miramonti and M.~D.~C.~Torri,
``Phenomenological Effects of CPT and Lorentz Invariance Violation in Particle and Astroparticle Physics,''
Symmetry \textbf{12} (2020) no.11, 1821
doi:10.3390/sym12111821
[arXiv:2110.09185 [hep-ph]].

\bibitem{Krauss}
L.~M.~Krauss and S.~Tremaine
''Test of the weak equivalence principle for neutrinos and photons,''
Phys. Rev. Lett. 1988 Jan 18;60(3):176-177
doi: 10.1103/PhysRevLett.60.176
PMID: 10038467

\bibitem{Gasperini}
M.~Gasperini
"Testing the principle of equivalence with neutrino oscillations"
Phys. Rev. D \textbf{38} (1988), 2635-2637
doi: 10.1103/PhysRevD.38.2635

\bibitem{Mann}
R.~B.~Mann and U.~Sarkar
"Test of the Equivalence Principle from Neutrino Oscillation Experiments"
Phys. Rev. Lett. \textbf{76} (1996), 865-868
doi: 10.1103/PhysRevLett.76.865

\bibitem{AmelinoFate}
G.~Amelino-Camelia,
Phys. Rev. D \textbf{85} (2012), 084034
doi:10.1103/PhysRevD.85.084034
[arXiv:1110.5081 [hep-th]].

\end{thebibliography}
\end{document}